\documentclass[aps,prf]{revtex4}
\usepackage{graphicx}
\usepackage{times}
\usepackage{color}
\usepackage{latexsym,amssymb,framed,amsbsy,amsmath}

\newcommand{\rem}[1]{}

\newcommand{\bel}{\begin{equation}\label}
\newcommand{\ee}{\end{equation}}
\newcommand{\beq}{\begin{eqnarray}} 
\newcommand{\eeq}{\end{eqnarray}} 
\newcommand{\bc}{\begin{center}} 
\newcommand{\ec}{\end{center}} 
\newcommand{\ben}{\begin{enumerate}}
\newcommand{\een}{\end{enumerate}}
\newcommand{\bit}{\begin{itemize}}
\newcommand{\eit}{\end{itemize}}

\newcommand{\bVq}{\mathcal{V}_{q}}
\newcommand{\bX}{\mathcal{X}}
\newcommand\shalf{\ensuremath{{\scriptstyle\frac{1}{2}}}}
\newcommand\sthird{\ensuremath{{\scriptstyle\frac{1}{3}}}}

\newcommand\sfourthirds{\ensuremath{{\scriptstyle\frac{4}{3}}}}
\newcommand{\bom}{\mbox{\boldmath$\omega$}}
\newcommand{\ba}{\mbox{\boldmath$a$}}
\newcommand{\bu}{\mbox{\boldmath$u$}}
\newcommand{\bv}{\mbox{\boldmath$v$}}

\newcommand{\bx}{\mbox{\boldmath$x$}}
\newcommand{\bdf}{\mbox{\boldmath$f$}}
\newcommand{\khat}{\mbox{\boldmath$\hat{k}$}}

\newcommand{\Rey}{{\rm Re}}
\newcommand{\Pec}{{\rm Pe}}
\newcommand{\Sc}{{\rm Sc}}
\newcommand{\Fr}{{\rm Fr}}

\begin{document}

\title{The Variable Density Model for the Rayleigh-Taylor Instability and its transformation to the diffusive, inhomogeneous, incompressible Navier-Stokes equations}

\author{John D. Gibbon} 
\affiliation{Department of Mathematics, Imperial College London, London SW7 2AZ, UK}

\begin{abstract}
It is shown how the variable density model (VDM) that governs the Rayleigh-Taylor instability (RTI) for the miscible mixing of two incompressible fluids can be transformed into a diffusive version of the inhomogeneous, incompressible Navier-Stokes equations forced by gradients of the composition density $\rho$ of the mixing layer. This demonstrates how buoyancy-driven flows drive and enhance Navier-Stokes turbulence. The role of the potential vorticity $q=\bom\cdot\nabla\rho$ is also discussed.
\end{abstract}

\date{\today} 
\keywords{Rayleigh-Taylor instability, Navier-Stokes equations}
\maketitle

\section{Introduction}\label{intro}

The Rayleigh-Taylor instability (RTI) is a phenomenon that occurs at the mixing interface of two fluids of different densities, initially set in a variety of  configurations. The literature documenting its occurrence is extensive, not only in fluid dynamics \cite{Sharp1984,Youngs1984,Youngs1989,BLMM2010,BLMM2012,BM2017,Liv2020,Glimm2001,Lee2008,Hyunsun2008,CPC2008} but also in astrophysics \cite{CabotCook2006}, plasma fusion \cite{Petrasso1994} and materials science  \cite{HohenbergHalperin1977,Bray1994,ChaikinLubensky2000}. The review by Dimotakis \cite{Dimotakis2005} discusses the Rayleigh-Taylor instability in the context of more general turbulent mixing processes. In fact, there are substantial differences within the Rayleigh-Taylor  process itself depending on whether the fluids are immiscible or miscible. In the immiscible case, various phase field models have been designed to overcome the severe challenges that arise in modelling the two-fluid interface \cite{CahnHilliard1958,Celani2009}. The miscible case has been widely explored by experiments in tanks, shock tubes, gas channels and other devices. For instance, in tank experiments the initial state is set up such that the heavier fluid sits over the lighter, separated by a barrier. On the removal of the barrier a turbulent mixing zone develops between the two which is supplied with kinetic energy by the conversion of potential energy stored in the initial configuration. This naturally drives a turbulent cascade down to small scales, with an increase in the dissipation rate of kinetic energy.  Such small scales also lead to substantially enhanced gradients in the density field which, in turn, also lead to irreversible mixing, and hence modification in the density distribution. More than a decade ago, Andrews and Dalziel wrote a review \cite{AndrewsDalziel2010} that focused on experiments at small Atwood numbers ($A_{t}\leq 0.1$)\,: subsequent work can be found in \cite{LawrieDalziel2011,Davies-WykesDalziel2014,Tailleux2013}. Banerjee's more recent review \cite{Banerjee2020}, written to commemorate the work and life of M. J. Andrews, has concentrated on RTI experiments in devices such as shock tubes and gas channels for which $A_{t}> 0.1$. 
\par\smallskip
There have been widely different approaches to modelling the behaviour of the mixing zone. One line of investigation has been to estimate the growth of the width of this zone, whose thickness has been observed to grow proportionately like $\alpha t^{2}$. Detailed measurements of $\alpha$ have been discussed by Banerjee, Kraft and Andrews \cite{BKA2010}. In terms of modelling the flow within this zone the variable density model (VDM), introduced by Sandoval \cite{Sandoval1995} and Cook and Dimotakis \cite{CookDimotakis2001}, is the most intriguing. Based on the Navier-Stokes equations, but with some subtle and important differences, the model is comprised of a set of equations that govern the evolution of the velocity field and composition density of the fluid in the mixing zone. Its properties have been explored computationally by Livescu and Ristorcelli \cite{LivescuRistorcelli2007,LivescuRistorcelli2008} and Aslangil, Livescu and Banerjee \cite{ALB2020}, and have been extensively summarized in the review by Livescu \cite{Liv2020}. The VDM considers two incompressible fluids of densities $\rho_{1}$ and $\rho_{2}$ whose mixing zone is considered to have a composition density $\rho(\bx,\,t)$ and a velocity field $\bu(\bx,\,t)$.  A brief derivation of this model is given in \S\ref{sec:2}. As a summary, it is emphasized there that the relation between the composition density $\rho$ and the velocity field $\bu$ is expressed in the form of a conservation-of-mass equation 
\beq\label{cmdef}
\partial_{t}\rho &+& \mbox{div}\left (\rho \bu \right) = 0\,,
\eeq
which, in turn, is moderated by an unusual equation for $\mbox{div}\,\bu$ \cite{CookDimotakis2001}
\beq\label{divudef}  
\mbox{div}\,\bu &=& - \Pec^{-1}\Delta (\ln\rho)\,. 
\eeq
The velocity field $\bu$ is then considered to satisfy the compressible Navier-Stokes equations with a gravitational forcing term $\Fr^{-2}\khat \rho$
\beq\label{fullcomNSE}
\rho\left(\partial_{t} + \bu\cdot\nabla\right)\bu &=& \Rey^{-1} \left\{\Delta\bu + \sthird\nabla\left(\mbox{div}\,\bu\right)\right\} - \nabla p + \Fr^{-2}\khat \rho\,.
\eeq
$\Pec$, $\Rey$ and $\Fr$ are respectively the P\'eclet, Reynolds and Froude numbers. The separate equation for $\mbox{div}\,\bu$ in (\ref{divudef}), with two derivatives of $\rho$, makes this an intriguing system of PDEs, quite unlike the standard model of incompressible Navier-Stokes fluids for which $\mbox{div}\,\bu=0$. The enforcement of this condition produces a Poisson equation for which the pressure $p$ can be determined. However, despite the fact that the mixing zone is varying in volume, the VDM is also unlike standard compressible flow. In that problem no separate equation for $\mbox{div}\,\bu$ exists, which necessitates an appeal to thermodynamic relations to express $p$ as a function of $\rho$, followed by some form of closure. As will be seen later in \S\ref{sec:3}, the existence of (\ref{divudef}) allows the pressure for the VDM system to be determined from a modified Poisson equation, although its solution presents certain difficulties which pushes the closure problem further down the line \cite{LivescuRistorcelli2007,LivescuRistorcelli2008,Liv2020}.
\par\smallskip
Having emphasised the unusual nature of the VDM system and its differences from both incompressible and compressible Navier-Stokes flows, 
it therefore comes as a surprise that the main result of this letter (see \S\ref{sec:3}) is that there exists an exact transformation between $(\bu,\,\rho)$ and a new velocity field $\bv$ defined by
\bel{vdef1}
\bv = \bu + \Pec^{-1}\nabla(\ln\rho)\,,
\ee
which transforms the three equations (\ref{cmdef}), (\ref{divudef}) and (\ref{fullcomNSE}) of the VDM into the simpler and more recognizable form of the \textit{diffusive, inhomogeneous, incompressible} Navier-Stokes equations \cite{Danchin2003,Danchin2004} expressed as 
\beq\label{iinse2a}
\left(\partial_{t} + \bv\cdot\nabla\right)\rho &=& \Pec^{-1}\Delta\rho\,,\\
\label{iinse2b}\mbox{div}\,\bv &=& 0\,,\\
\label{iinse2c}
\rho\left(\partial_{t} + \bv\cdot\nabla\right)\bv &=&  \Rey^{-1} \Delta\bv - \nabla \tilde{p} + \rho\bdf\,, 
\eeq
where $\tilde{p}$ is a modified pressure and $\bdf$ is the sum of a constant and the gradient of a function of $\rho$ and up to two of its derivatives. Results and data sets for (\ref{iinse2a}) -- (\ref{iinse2c}) can thus be mapped to the VDM. 
\par\smallskip
Finally, in the context of GFD mixing processes, the evolution of the potential vorticity $q=\bom\cdot\nabla\rho$ is discussed in \S\ref{sec:4}.

\section{Brief derivation of the Variable Density Model (VDM)}\label{sec:2}

Consider two incompressible, miscible fluids with constant densities $\rho_{1} < \rho_{2}$. Despite the fact that the two fluids are themselves incompressible, molecular mixing generically changes the specific volume of the mixture. This type of flow is called a \textit{variable density} flow, following the nomenclature suggested by Cook and Dimotakis \cite{CookDimotakis2001}, and Livescu and Ristorcelli \cite{LivescuRistorcelli2007,LivescuRistorcelli2008,Liv2020}.  In variable density flows, because the specific volume of the mixture is not constant, it is necessary to define what is called the \textit{composition density} $\rho(\bx,\,t)$ of a mixture of two constant fluid densities $\rho_{1}$ and $\rho_{2}$ which, in dimensionless form, is defined to be
\bel{sma1}
\frac{1}{\rho(\bx,\,t)} = \frac{Y_1(\bx,\,t)}{\rho_{1}} + \frac{Y_2(\bx,\,t)}{\rho_{2}}\,,
\ee
where $Y_{i}(\bx,\,t) > 0$ ($i=1,2$) are the mass fractions of the two fluids subject to $Y_1+Y_2=1$. (\ref{sma1}) shows that $\rho$ is bounded by $\rho_{1} < \rho(\bx,\,t) < \rho_{2}$. Let us write this in the more general $N$-component form
\bel{sma2}
\frac{1}{\rho(\bx,\,t)} = \sum_{i=0}^{N}\frac{Y_{i}(\bx,\,t)}{\rho_{i}}\qquad\mbox{subject~to}\qquad \sum_{i=0}^{N}Y_{i} = 1\,,
\ee
where specifically for this problem $N=2$. To determine how (\ref{sma2}) couples to a corresponding velocity field $\bu(\bx,\,t)$, it is assumed that there is Fickian diffusion. Then the mass transport equation for each of these components is given by
\bel{sma3} 
\partial_{t}\left( \rho Y_{i}\right) + \mbox{div}\left(\rho Y_{i}\bu \right) = 
\Pec^{-1}\mbox{div}\left(\rho\nabla Y_{i}\right)\,,
\ee
where $\Pec$ is the P\'eclet number. The Reynolds ($\Rey$) and P\'eclet ($\Pec$) numbers are related by $\Pec = \Rey\Sc$ where $\Sc$ is the Schmidt number. The conventional continuity equation for mass conservation,
\bel{cm1}
\partial_{t}\rho + \mbox{div}\left(\rho \bu \right) = 0\,,
\ee
is simply derived by taking the sum of (\ref{sma3}) and using the fact that the $Y_{i}$ sum to unity as in (\ref{sma2}). Next, we note that because (\ref{sma3}) is true for each value of $i$, we divide by $\rho_{i}$, sum over $i$ and use (\ref{sma2}). After minor manipulation, the final result simplifies to
\bel{cm2}
\mbox{div}\,\bu = - \Pec^{-1}\Delta(\ln\rho)\,.
\ee
The velocity field $\bu$ is assumed to obey the compressible Navier-Stokes momentum equation with the effect of gravity included
\beq\label{cm3}
\rho\left(\partial_{t} + \bu\cdot\nabla\right)\bu &=& \Rey^{-1} \left\{\Delta\bu + \sthird\nabla\left(\mbox{div}\,\bu\right)\right\} - \nabla p + \Fr^{-2}\khat \rho\,.\label{vde3}
\eeq
The PDEs (\ref{cm1}), (\ref{cm2}) and (\ref{cm3}) constitute the VD-model. It is worth remarking as an aside that this result is true for all values of $N\geq 2$, which would constitute a more general mixing problem than the specific two-component RT-model investigated here. Livescu and Ristorcelli  \cite{Liv2020,LivescuRistorcelli2007,LivescuRistorcelli2008} have performed extensive computations using periodic boundary conditions at small Atwood number
\bel{Atdef}
A_{t} = \frac{\rho_{2} - \rho_{1}}{\rho_{2} + \rho_{1}} \sim 0.1\,,
\ee
while Aslangil, Livescu and  Banerjee \cite{ALB2020} have considered different regimes of $A_{t}$ and $\Rey$.

\section{A transformation to the diffusive inhomogeneous, incompressible Navier-Stokes equations}\label{sec:3}

The purpose of this section is to show that with the definition $\theta = \ln \rho$, a new velocity field defined as
\bel{vdef}
\bv = \bu + \Pec^{-1}\nabla\theta\,,\qquad\mbox{where}\qquad \mbox{div}\,\bv = 0\,,
\ee
transforms the set of equations (\ref{cm1})--(\ref{cm3}) (or equivalently (\ref{cmdef})--(\ref{fullcomNSE})) into (\ref{iinse2a}), (\ref{iinse2b}) and (\ref{iinse2c}), where $\bdf$ and $\tilde{p}$ are defined by
\beq\label{bdf3a}
\bdf &=& \nabla\phi + \Fr^{-2}\khat\,;\qquad
\tilde{p} = p + \psi\\
\label{bdf3b}\psi &=& \sfourthirds \Rey^{-1}\Pec^{-1}\Delta\theta\,\\
\label{bdf3c}\phi &=& \Pec^{-2}\left\{\Delta\theta + \shalf(\nabla\theta)^{2}\right\}\,.
\eeq
\par\smallskip\noindent
Before embarking on the proof, a remark is in order. The transformation in (\ref{vdef}) produces the inhomogeneous, incompressible Navier-Stokes equations, with diffusion in $\rho$, but with the effect of $\nabla\theta$ and $\Delta\theta$ appearing in $\phi$ within the forcing in the $\rho\nabla\phi$-term. In other words, $\rho$ diffuses in the standard way but turbulent intermittency is driven by gradients of $\theta$.  A study of the data confirms that $\nabla\theta$ can become very large \cite{RCG2017}. 
\par\smallskip
Using the new velocity field defined in (\ref{vdef}), it is easy to see that (\ref{cm1}) transforms to
\bel{vdrho}
\left(\partial_{t} + \bv\cdot\nabla\right)\rho = \Pec^{-1}\Delta\rho\,.
\ee
Now let us define a `new' material derivative such that 
\bel{Ddef}
\frac{D~}{Dt} = \partial_{t} + \bv\cdot\nabla\,.
\ee
Then the `old' material derivative $\left(\partial_{t} + \bu\cdot\nabla\right)$ acting on $\bu$ is
\beq\label{nse4}
\left(\partial_{t} + \bu\cdot\nabla\right)\bu &=& \frac{D~}{Dt}\left(\bv -  \Pec^{-1}\nabla\theta\right) - \Pec^{-1}\nabla\theta
\cdot\nabla\left(\bv -  \Pec^{-1}\nabla\theta\right)\nonumber\\
&=& \frac{D\bv}{Dt} - \Pec^{-1}\nabla\left(\frac{D\theta}{Dt}\right) + \Pec^{-2}\nabla\left(\shalf|\nabla\theta|^{2}\right)\,.
\eeq
Note that the last term on the last line appears because of the vector identity $\shalf\nabla(\ba\cdot\ba) = \ba\cdot\nabla\ba + 
\ba\times\mbox{curl}\ba$, so when $\ba$ is a gradient, the curl- term is zero. Thus (\ref{vde3}) becomes 
\beq\label{nse5}
\rho\frac{D\bv}{Dt} &-& \Pec^{-1}\rho\left\{\nabla\left(\frac{D\theta}{Dt}\right) - \Pec^{-1}\nabla\left(\shalf|\nabla\theta|^{2}\right)\right\}\\
&=& - \nabla p + \Rey^{-1} \Delta\left(\bv - \Pec^{-1}\nabla\theta\right)
- \sthird \Rey^{-1} \Pec^{-1}\nabla\left(\Delta\theta\right) + \Fr^{-2}\khat \rho\,,\nonumber
\eeq
which can be rewritten as
\beq\label{nse6}
\rho\frac{D\bv}{Dt} + \nabla \left(p + \sfourthirds \Rey^{-1}\Pec^{-1}\Delta \theta \right) - \Rey^{-1}\Delta\bv &=& 
\Pec^{-1}\rho\nabla\left\{\frac{D\theta}{Dt} - \shalf\Pec^{-1}|\nabla\theta|^{2}\right\} + \Fr^{-2}\khat \rho\,.
\eeq
Noting that $D\theta/Dt = \Pec^{-1}\left(\Delta\theta +|\nabla\theta|^{2}\right)$, we have 
\beq\label{nse7}
\rho\frac{D\bv}{Dt}  + \nabla \left(p + \sfourthirds \Rey^{-1}\Pec^{-1}\Delta \theta \right) - \Rey^{-1} \Delta\bv &=& \Pec^{-2}\rho\nabla
\left\{\Delta\theta + \shalf |\nabla\theta|^{2}\right\} + \Fr^{-2}\khat \rho\,,
\eeq
which is (\ref{iinse2c}) together with (\ref{bdf3a})--(\ref{bdf3c}). 
\par\smallskip
Finally, we remark that $\psi$ and $\phi$ in (\ref{bdf3b}) and (\ref{bdf3c}) can be expressed as 
\beq\label{bdf3d}
\psi &=& \sfourthirds \Rey^{-1}\Pec^{-1}\rho^{-2}\left\{\rho\Delta\rho - (\nabla\rho)^{2}\right\}\\
\label{bdf3e}\phi &=& \Pec^{-2}\rho^{-2}\left\{\rho\Delta\rho - \shalf(\nabla\rho)^{2}\right\}\,.
\eeq
which shows that the system decouples when these are negligible. It has been observed that $|\nabla\rho|$ can become very large \cite{RCG2017}, thus making $\bdf$ a term that heavily forces Navier-Stokes turbulence. 
\par\smallskip
Because of the multiplicative factor of $\rho$ in (\ref{iinse2c}) the application of the constraint $\mbox{div}\,\bv = 0$ across (\ref{iinse2c}) produces a Poisson equation for $\tilde{p}$ which has extra terms, including one in $\nabla\tilde{p}$, not present in the standard incompressible Navier-Stokes Poisson equation for the pressure 
\bel{poiss2}
\Delta\tilde{p} + \nabla(\ln\rho)\cdot \left(\Rey^{-1} \Delta\bv - \nabla \tilde{p}\right) =  - \rho\sum_{i,j}v_{i,j}v_{j,i} + \rho\,\mbox{div}\bdf\,.
\ee
The difficulties thrown up by these extra terms, whose origin lies in (\ref{cm2}), have been discussed at length in \cite{LivescuRistorcelli2007,LivescuRistorcelli2008,Liv2020}.

\section{A remark on the potential vorticity}\label{sec:4}

In geophysical fluid dynamics (GFD) the potential vorticity $q = \bom\cdot\nabla\rho$ is considered a key quantity in understanding cyclogenesis, so its behaviour is worth examining in the context of VDM mixing processes \cite{Dimotakis2005}.  In the Euler limit, $q$ is a material constant when the density is also a material constant. With the presence of viscosity we are able to discuss how close $q = \bom\cdot\nabla\rho$ and the composition density $\rho$ come to being material constants. This is based on an idea that can be found in \cite{GibbonHolm2010,HM1990}.
\par\medskip
Firstly, we note that $\rho$ has no zeros as it is bounded below by $\rho_{1}$. Secondly, the vorticity $\bom = \mbox{curl}\,\bv = \mbox{curl}\,\bu$ satisfies 
\bel{qp1}
\frac{D\bom}{Dt} = \Rey^{-1}\rho^{-1}\Delta\bom + \Rey^{-1}\nabla\left(\rho^{-1}\right)\times\Delta\bv + \bom\cdot\nabla\bv
- \nabla\left(\rho^{-1}\right)\times\nabla\tilde{p}\,.
\ee
The forcing $\bdf$ has disappeared under the curl-operation because it is the sum of a constant and a gradient function. Then we have 
\beq\label{qp2}
\frac{Dq}{Dt} &=& \left\{\Rey^{-1}\rho^{-1}\Delta\bom + \Rey^{-1}\nabla\left(\rho^{-1}\right)\times\Delta\bv + \bom\cdot\nabla\bv - \nabla\left(\rho^{-1}\right)\times\nabla\tilde{p}\right\}\cdot\nabla\rho\nonumber\\
&+& \bom\cdot\left\{\nabla\left(\frac{D\rho}{Dt}\right) - (\nabla\bv)\cdot\nabla\rho\right\}
\eeq
Four terms disappear within (\ref{qp2}), two by cancellation \cite{Ertel1942,GibbonHolm2010}, and two as scalar triple products, leaving 
\beq\label{qp3}
\frac{Dq}{Dt} &=& \Rey^{-1}\Delta\bom\cdot\nabla(\ln\rho) + \Pec^{-1}\bom\cdot\nabla\Delta\rho\nonumber\\
&=& -\mbox{div}\left\{\Rey^{-1}\nabla(\ln\rho)\times\Delta\bv - \Pec^{-1}\bom\Delta\rho \right\}
\eeq
Thus we obtain 
\bel{qp4}
\partial_{t}q + \mbox{div}\left\{q\bv + \Rey^{-1}\nabla(\ln\rho)\times\Delta\bv - \Pec^{-1}\bom\Delta\rho \right\} = 0\,.
\ee
Then $q$ and $\rho$ can be seen to obey
\bel{q1}
\partial_{t}q + \mbox{div}\,(q\bVq) = 0\,,\qquad\qquad \partial_{t}\rho + \bVq\cdot\nabla\rho = 0\,,
\ee
where
\bel{q2}
q\left(\bVq - \bv\right) = \bX_{q}\,,\qquad\mbox{and}\qquad \bX_{q} = \Rey^{-1}(\nabla(\ln\rho)\times\Delta\bv) - \Pec^{-1}\bom\Delta\rho\,.
\ee
Equations (\ref{q1}) and (\ref{q2}) are a simple formulation of the problem at the level of the vorticity which does not directly involve the pressure. The price one pays for its absence is the fact that the formulation breaks down when $q=0$ and that $\mbox{div}\,\bVq \neq 0$. This raises two questions\,: i) how often in a numerical computation does $q$ change sign, and ii) what is the value of $\mbox{div}\,\left(q^{-1}\bX_{q}\right)$?  It is the value of the latter that shows how far $\mbox{div}\,\bVq$ lies from zero and thus prevents both $q$ and $\rho$ from being exact material constants.

\vspace{-3mm}
\section{Conclusion}\label{sec:5}

The results in this letter show that the equations governing the VD-model can be transformed into the diffusive, inhomogeneous, incompressible Navier-Stokes equations with a forcing $\bdf$ which is a constant plus a gradient function of up to two derivatives of $\theta$. Thus, we can argue that Rayleigh-Taylor turbulence is indeed Navier-Stokes turbulence driven by gradients of the buoyancy. This is consistent with evidence that buoyancy effects are particularly efficacious in forcing turbulent flow 
\cite{Dimotakis2005,LawrieDalziel2011,Davies-WykesDalziel2014,Tailleux2013,CPC2021}. Nevertheless, while the inhomogeneous, incompressible Navier-Stokes equations are a much simpler system, analytically they remain an order of magnitude more difficult to handle than the fully incompressible Navier-Stokes equations \cite{Danchin2003,Danchin2004}. Moreover, computations are still plagued by difficulties in solving for the pressure \cite{Liv2020,LivescuRistorcelli2007,LivescuRistorcelli2008}.


\end{document}